\documentclass[twocolumn,showpacs,preprintnumbers,amsmath,amssymb]{revtex4}
\usepackage{epsfig}
\usepackage{amsmath}

\begin{document}

\title{Entropy Moments Characterization of Statistical Distributions}

\author{Luciano da Fontoura Costa}
\affiliation{Institute of Physics at S\~ao Carlos, University of
S\~ao Paulo, P.O. Box 369, S\~ao Carlos, S\~ao Paulo, 13560-970
Brazil}

\date{21st March 2008}

\begin{abstract}
This letter reports two moment extensions of the entropy of a
distribution.  By understanding the traditional entropy as the average
of the original distribution up to a random variable transformation,
the traditional moments equation become immediately applicable to
entropy.  We also suggest an alternative family of entropy moments.
The discriminative potential of such entropy moment extensions is
illustrated with respect to different types of distributions with
otherwise undistinguishable traditional entropies.
\end{abstract}

\maketitle

\vspace{0.5cm}
\emph{`Looking into each globe, you see a blue city, the model of
a different Fedora.'
(I. Calvino, Invisible Cities)}
\vspace{0.5cm}

Given a statistical distribution $p(x)$, where $x$ is the respective
random variable, an important problem is to try to synthetize its most
important features into as few measurements $f_i$, $i = 1,2, \ldots,
N$ as possible.  Although a distribution incorporates all information
about the respective random variable $x$, it typically involves a
large number of values.  While continuous distributions have infinite
values, discrete distributions typically involve a large number of
bins.  However, the summarization of a distribution in terms of a few
respective functionals is not straightforward and ultimately depends
on specific goals.  For instance, one may be interested in intervals
of regularity along the distributions, or in the overall dispersion.
Generally, it is useful to remove the redundancy from the
distributions, leaving out only the most informative variations and
singularities.  Traditional functionals of distributions include the
respective moments given as

\begin{equation}
  M(p(x),k) = \int_I x^k p(x) dx  \label{eq:mom}  
\end{equation}

where $I$ is the domain of $p(x)$, i.e. its sampling space. Observe
that these moments have the same dimensionality as the original random
variable $x$.  The first moment corresponds to the average and the
second moment is related to the variance of the random variable $x$.
It is know from statistical theory (e.g.~\cite{Shohat:1943,
Dudewicz:1988}) that the set of all infinite moments can, under
certain conditions (the so-called moment problem), provide a complete
mapping of the original distribution, in the sense that the latter can
be recovered from the former.  Generally, increasing information about
the features of the original distribution can be obtained by
considering a larger number of moments.  Another important functional
of a statistical distribution is its respective \emph{entropy}
(e.g.~\cite{Greven:2003, Sethna:2006, Cover:1991}), which is defined
as

\begin{equation}
  e(p(x)) = -\int_I p(x) log(p(x)) dx
\end{equation}

This measurement becomes zero for distributions involving identical
values of $x$, being maximized for uniform distributions,
i.e. identical values of $p(x)$ along $I$.  The entropy measurement
exhibits several particularly relevant properties, including its
intrinsic relationship with statistical physics
(e.g.~\cite{Landau:1980}), entropy maximization (e.g.~\cite{Boyd:2005,
Cover:1991}), information theory and channel capacity
(e.g.~\cite{Cover:1991}).  The entropy is also invariant to
transformation of the values of $x$, i.e. the entropy of the
distribution of $x$ is identical to the distribution of the new random
variable $y=f(x)$, where $f$ is any one-to-one function.  Yet,
typically the entropy is considered as an isolated measurement.

In this article we suggest a family of entropy-based measurements
which provide enhanced information about the original distribution.
First, we show that the entropy can be understood as a special case of
the first moment, where the values of the random variable $x$ are
substituted by the adimensional quantity $log(p(x))$, i.e. the weights
in the average definition are exchangec by the logarithm of the
distribution values.  By doing so, it becomes possible to calculate
all respective moments and central moments, which are henceforth
called the \emph{entropy moments} and \emph{entropy central moments}.
We illustrate the power of such additional statistical measurements
with respect to the discrimination between important types of
statistical distributions.

We henceforth focus our attention on discrete distributions
represented in the continouous space of the variable $x$, i.e.

\begin{equation}
  p(x) = p(x_i) = \sum_{i=1}^{N} p(i) \delta(x_i)
\end{equation}

where $x$ is a continuous variable in $I = [a,b]$ and $\delta(x_i)$ is
the Dirac's delta function placed at $x_i$, i.e. $\delta_(x_i) =
\delta_{x-x_i}$, with $i = 1, 2, \ldots, N$.  Therefore, $p(x_i)$ can
be used to represent any relative frequency histogram. The moments of
this distribution are immediately given by Equation~\ref{eq:mom}.

Now, by introducing the new random variable $y_i = log(p(x_i))$, we
can rewrite the entropy as

\begin{equation} 
  e(p(x_i)) = -\int_I p(x_i) log(p(x_i)) dx 
            = -\int_J y_i p(y_i) dy   \label{eq:new}
\end{equation}

where $J$ is the mapped version of the interval $I$, i.e. $J =
[min(log(p(x_i))),max(log(p(x_i)))]$.  Observe that $p(x_i) = p(y_i)$
for any $i = 1, 2, \ldots, N$.  We have from Equation~\ref{eq:new}
that the traditional entropy can be understood as the negative of the
first moment (i.e. average) of the distribution of the transformed
random variable $y_i = log(p(x_i))$.  The extension to higher order
moments is straightforward and yields the respective moments given by
Equation~\ref{eq:ME}.

\begin{eqnarray}
  \mathrm{ME}(p(x_i),k) = - \int_J (y_i)^k p(y_i) dy   \label{eq:ME}  \\
  \mathrm{FE}(p(x_i),k) = log \left(- 
                          \int_J (p(y_i))^k y_i dy \right)  \label{eq:FE} 
\end{eqnarray}

We necessarily have that $\mathrm{ME}(p(x_i),k) =
\mathrm{ME}(p(y_i),k)$ and $\mathrm{FE}(p(x_i),k) =
\mathrm{FE}(p(y_i),k)$.  Observe that the non-dimensionality of
$y_i$ is immediately extended to the entropy moments.  In addition,
the consideration of $y_i$ as the weights for the moment calculation
implies the respectively induced distribution $p(y_i)$ to be sorted
into ascending order. It should be also observe that the successive
entropy moments $ME$ tend to present inverse signals.  Because of the
moment mapping theorem, we have that all the information in the
original distribution $p(x_i)$ is captured by the infinite set of
respective moments.  Therefore, these additional entropy-based
measurements provide an interesting complementation of the traditional
entropy, allowing a more comprehensive characterization of the
original distribution in terms of a set of respective functionals, in
direct analogy with the role of the traditional moments.  The
alternative entropy moments defined by Equation~\ref{eq:FE} have been
found to allow particularly discriminating measurements.  In this
definition, the most external logarithm is used in order to obtain
more manageable values.

\begin{figure*}[htb]
  \vspace{0.3cm} 
  \begin{center}
  \includegraphics[width=1\linewidth]{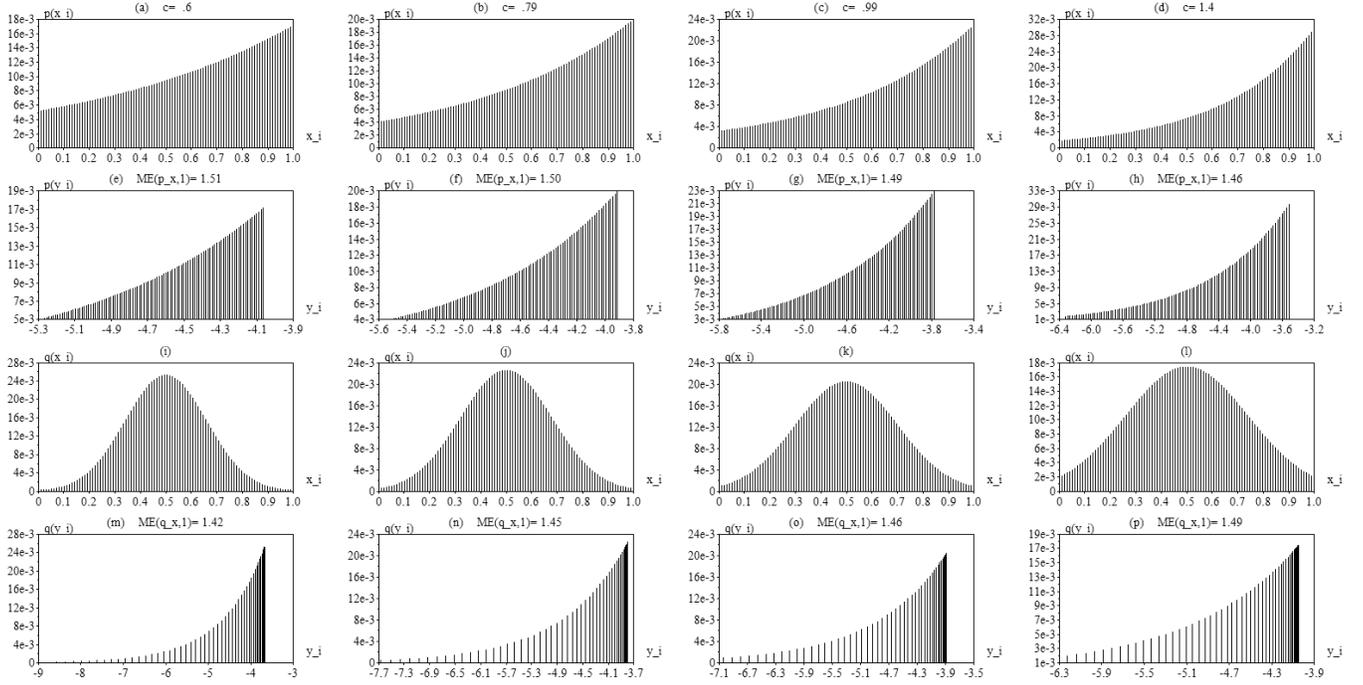} \\  
  \caption{The distributions $p(x_i) = w\ exp(ci)$ (a-d) and 
             $q(x_i)=s\ exp(-0.5((ci-\mu)/\sigma)^2)$ (i-l)
             as well as their respective
             transformations $p(y_i)$  and $q(y_i$ for several values of $c$.
  }~\label{fig:dists} 
  \end{center}
\end{figure*}

In the remainder of this article, we provide a series of examples of
the potential of the entropy moments and alternative entropy moments.
First, we consider distributions of the type $p(x_i) = w\ exp(ci)$,
$x_i = (i-1)\Delta$ and $\Delta = (b-a)/(N-1)$,where $c$ is a real
value such that $0 \leq c$ and $w$ is a normalizing constant ensuring
$\int_{I} p(x_i) dx = 1$.  Observe that this distribution becomes the
uniform distribution when $c=0$ and the constant distribution when $c
\rightarrow \infty$. Figure~\ref{fig:dists} illustrates the
distribution $p(x_i)=w\ exp(ci)$ (a-d) and the normal distribution
$q(x_i)=s\ exp(-0.5((ci-\mu)/\sigma)^2)$ (i-l), where $s$ is a
normalizing constant, as well as the respective transformed
distributions $p(y_i)$ (e-h) and $q(y_i)$ (m-p) for several values of
$c$, assuming the values of $x_i$ to be distributed at equal spaces
along $I = [0,1]$.  Observe that the distribution $p(x_i)$ tends to
become less uniform for larger values of $c$ (moving from
Fig.~\ref{fig:dists}a to d), while the opposite is verified for
$q(x_i)$ (moving from Fig.~\ref{fig:dists}i to l).  Such trends are
clearly reflected in the respective entropy values (i.e. $e(p_x) =
ME(p(x_i),1)$) shown above the respective transformed distributions in
Figure~\ref{fig:dists}(e-h) and (m-p), respectively.  It is also clear
from Figure~\ref{fig:dists}, particularly for the distribution
$q(x_i)$, that the transformed distribution $p(y_i)$ is sorted in
increasing order as a consequence of the random variable
transformation $y_i = log(p(x_i))$.  Observe also the increased
density of Dirac's deltas at the right-hand side of the distributions
in Figure~\ref{fig:dists}(m-p), which are a consequence of the similar
values of the normal distribution $q(x_i)$ near its peak.

\begin{figure}[htb]
  \vspace{0.3cm} \begin{center}
  \includegraphics[width=0.8\linewidth]{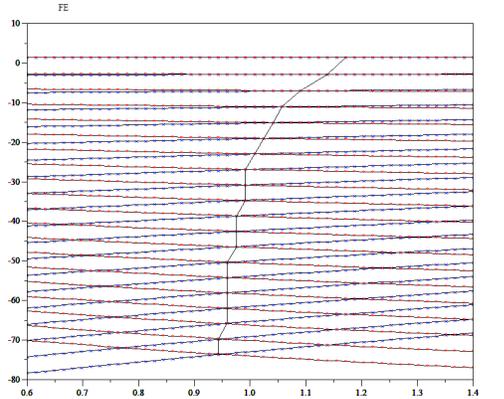} \\ 
  \caption{
  The entropy moments of $p(x_i)$ (red) and $q(x_i)$ (blue) in
  terms of $c$ for several values of $k$ (shown from top to bottom).  
  The firsr upper red and blue upper curves (above 0) correspond 
  to the traditional entropies of $p(x_i)$ and $q(x_i)$.
  }~\label{fig:ents}
  \end{center}
\end{figure}

Figure~\ref{fig:ents} depicts the set of alternative moment entropies
$\mathrm{FE}(p(x_i),k)$ of the distributions $p(x_i)$ (a) and $q(x_i)$
(b) as above, in terms of $c$ for several values of $k$, i.e. the
order of the alternative entropy moments.  The points where the
alternative entropy moments of $p(x_i)$ and $q(x_i)$ equal one another
have been marked by the `vertical' trajectory.  It is clear from these
results that though the distributions $p$ and $q$ have identical
traditional entropy for $c \approx 1.06$, substantial differences are
observed between the higher order alternative entropy moments.  
Interestingly, though the first alternative entropy moment (identical
to the traditional entropy) increases with $c$ as expected, the higher
order moments tend to decrease with $c$.

\begin{figure}[htb]
  \vspace{0.3cm} 
  \begin{center}
  \includegraphics[width=0.95\linewidth]{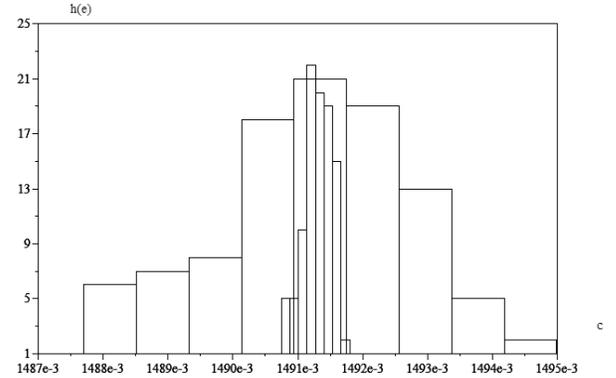} \\  
  \caption{The histograms of the traditional entropy obtained for 
             the perturbed versions of the two
             distributions $p(x_i)$ and $q(x_i)$.  Because of the
             complete overlat between these two histograms, it
             is completely impossible to discriminate between the
             original distributions while considering their 
             respective traditional entropies.
  }~\label{fig:hist} 
  \end{center}
\end{figure}

In order to better illustrate the potential of the entropy moments for
providing additional information about the original distribution, we
now focus our attention on the two above distributions $p(x_i)$ and
$q(x_i)$ at a value of the parameter $c$ at which they can by no means
be discriminated by considering the respective traditional entropies.
In order to simulate sampling noise and artifact typically implied
while measuring the random variable $x$, we add a uniformly
distributed perturbation to each of the two distributions.
Figure~\ref{fig:hist} shows the histograms of the traditional entropies
calculated for the two perturbed distributions.  Because of the
complete superposition between the respective histograms, it is
virtually impossible to discriminate between the two cases while
taking into account their respective traditional entropies.

\begin{figure*}[htb]
  \vspace{0.3cm} 
  \begin{center}
  \includegraphics[width=1\linewidth]{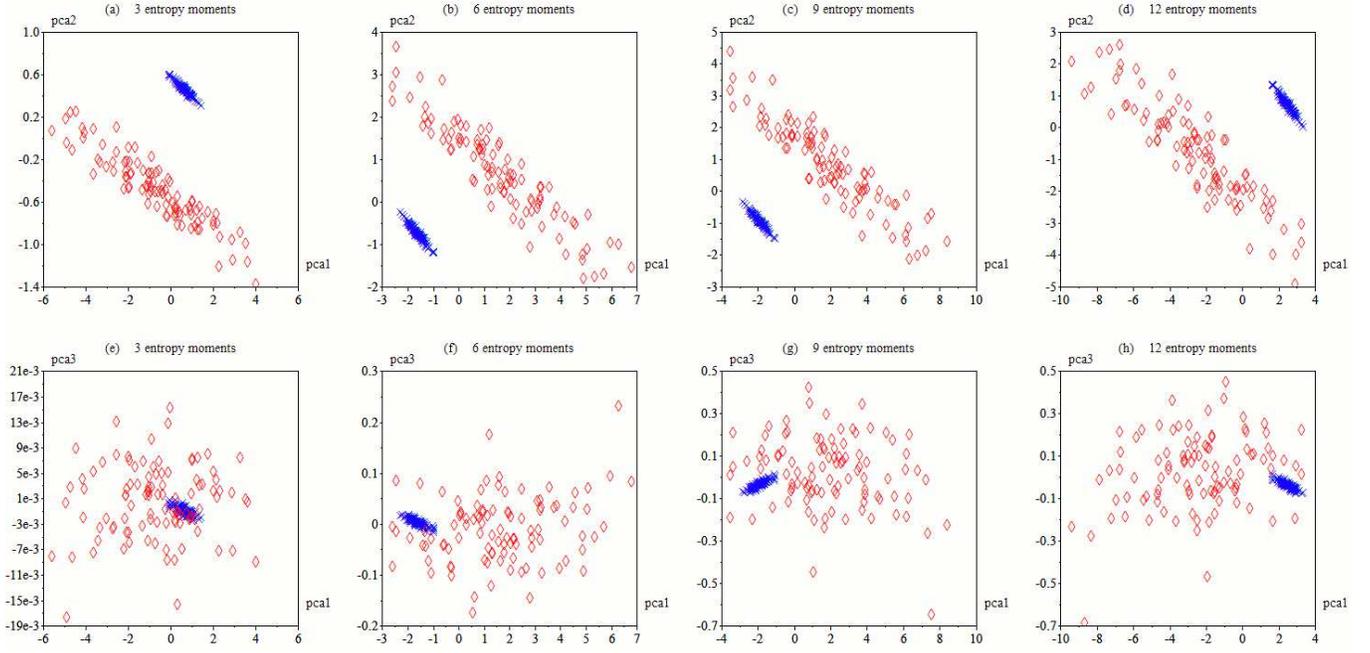} \\  
  \caption{ The scattering of the two categories of perturbed distributions
            as revealed by two-dimensional projection (through PCA) of 
            respective characterizations incorporating 3 (a), 6 (b), 9 (c)     
            and 12 (d) entropy moments.
  }~\label{fig:pca_2} 
  \end{center}
\end{figure*}

We now consider the effect of the consideration of additional entropy
moments on the discriminability between the measurements.
Figure~\ref{fig:pca_2} illustrates the scattering of the entropy
moments obtained for the perturbed realizations of the two types of
distributions (i.e. $p_(x_i)$ and $q(x_i)$) considering 3 (a), 6 (b),
9 (c) and 12 (d) entropy moments.  The two-dimensional projections
shown in Figure~\ref{fig:pca_2} were obtained by using the principal
component analysis (PCA) methodology (e.g.~\cite{Duda_Hart:2001,
Fukunaga:1990, Costa_book:2001}), which ensures maximum dispersion
along the first axes of the projections, which are defined by the
transformed variables $pca_i$, $i=1, 2, \ldots$.  More specifically,
the PCA involves the calculation of the covariance matrix of the
considered measurements and estimation of the respective eigenvalues
and eigenvectors.  The linear transformation used to project the
higher dimensional space is defined by the eigenvectors of the
covariance matrix taken in decreasing order.  In order to compensate
for the largely different values of the entropy moments, their values
were standardized~\footnote{The standardization of a random variable
involves subtraction by the average and division by the standard
deviation (e.g.~\cite{Duda_Hart:2001, Costa_book:2001}).  The values
of the transformed random value tends to be comprised between -2 and
2.} prior to the PCA.  It is clear from the results shown in
Figure~\ref{fig:pca_2}(a-d) that the incorporation of additional
entropy moments contributed substantially for the separation between
the perturbed cases.  However, the consideration of additional entropy
moments tended not to enhance such a separation.  For instance, the
separation between the two perturbed distributions considering 3
entropy moments (Fig.~\ref{fig:pca_2}a) is similar to that obtained
for 12 entropy moments (Fig.~\ref{fig:pca_2}d).  In addition, the
contribution of the higher order entropy moments had almost no effect
in increasing the separation between the two categories of
observations while considering the third principal component axis,
i.e. $pca3$ (see Figs.~\ref{fig:pca_2}e-h).

\begin{figure*}[htb]
  \vspace{0.3cm} 
  \begin{center}
  \includegraphics[width=1\linewidth]{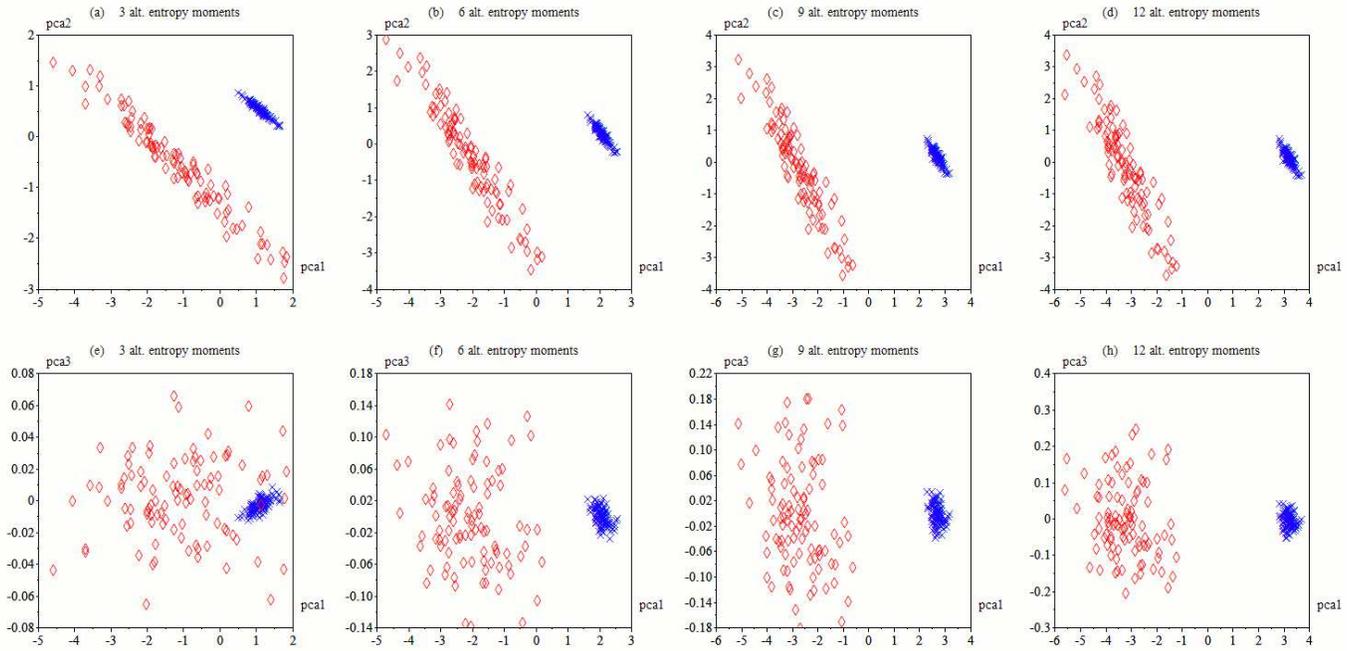} \\  
  \caption{ The scattering of the two categories of perturbed distributions
            as revealed by two-dimensional projection (through PCA) of 
            respective characterizations incorporating 3 (a), 6 (b), 9 (c)     
            and 12 (d) alternative entropy moments.
  }~\label{fig:pca} 
  \end{center}
\end{figure*}

Figure~\ref{fig:pca} shows the PCA results considering alternative
entropy moments, instead of the entropy moments as above.  The
incorporation of additional alternative entropy moments allows the
increasing discrimination between the two sets of observations
regarding all the three first PCA variables (i.e. $pca1$, $pca2$ and
$pca3$).

All in all, we have reported on two families of entropy moments,
obtained by interpreting the traditional entropy as the average of a
transformed version of the original distribution.  Such additional
measurements have been shown to contribute substantially for the
characterization of the original distributions, as clearly illustrated
for a case involving two distributions with undistinguishable
traditional entropies.  Because of the key role played by entropy in
so many areas, the concepts and results described in this work have
several immediate implications. Among the several possibilities for
future developments, we have the investigation of entropy central
moments, including the development of a PCA methodology based on the
respectively implied entropy covariance matrix.  It would also be
interesting to investigate the type of distribution features which
lead to extreme values of each of the entropy moments.

\begin{acknowledgments}
Luciano da F. Costa thanks CNPq (301303/2006-1) and FAPESP (05/00587-5)
for sponsorship.
\end{acknowledgments}

\bibliography{altent}
\end{document}